\documentclass{aa}
\usepackage{graphicx}
\usepackage{natbib}
\usepackage{amsmath}
\usepackage{txfonts}
\begin{document}

\title{Soft X-ray components in the hard state of accreting black holes}
\titlerunning{Soft X-ray components in the hard state of accreting black holes}
\authorrunning{D'Angelo et al.}
\author{Caroline D'Angelo\inst{1}\thanks{dangelo@mpa-garching.mpg.de}
\and Dimitrios Giannios\inst{1}
\and Cornelis Dullemond\inst{2}
\and Henk Spruit\inst{1}}

\institute{Max-Planck-Institut f\"ur Astrophysik,
  Karl-Schwarzchild-Str. 1, 85741 Garching, Germany 
\and Max-Planck-Institut f\"ur Astronomie, K\"onistuhl 17, 69177
Heidelberg, Germany}

\date{Received / Accepted}

\abstract{Recent observations of two black hole candidates (GX
339-4 and J1753.5-0127) in the low-hard state
($L_{\rm{X}}/L_{\rm{Edd}} \simeq 0.003-0.05$) suggest the presence of
a cool accretion disk very close to the innermost stable orbit of the
black hole. This runs counter to models of the low-hard state in which
the cool disk is truncated at a much larger radius. We study the
interaction between a moderately truncated disk and a hot inner
flow. Ion-bombardment heats the surface of the disk in the overlap
 region between a two-temperature advection-dominated accretion
flow and a standard accretion disk, producing a hot
($kT_{\rm{e}}\simeq 70$keV) layer on the surface of the cool disk. The
hard X-ray flux from this layer heats the inner parts of the
underlying cool disk, producing a soft X-ray excess. Together with
interstellar absorption these effects mimic the thermal spectrum from
a disk extending to the last stable orbit. The results show that soft
excesses in the low-hard state are a natural feature of truncated disk
models.
}
\maketitle

\section{Introduction}
The geometry of the low-luminosity (``low-hard'') state of Galactic
Black Hole Candidates (GBHC), in which the spectrum is dominated by a
power law X-ray flux extending to high energies, has been an open
question for several decades. While it is generally believed that the
power law spectrum is formed by inverse Compton scattering, there is
no consensus about the geometry of the flow, source of seed photons or
energy distribution for the Comptonizing electrons.

 Broadly speaking, there are two classes of model to explain the
spectrum in the low-hard state. The first is the ``corona'' model, in
which the disk remains untruncated or nearly untruncated at
luminosities $L_{\rm{X}} \simeq 10^{-3} L_{\rm{Edd}}$. The hard power
law spectrum comes from a hot and patchy corona (perhaps powered by
magnetic flares \citep{1998MNRAS.299L..15D, 1999ApJ...510L.123B,
2001MNRAS.321..549M}) on top of the disk, while the surrounding region
is bombarded with high energy photons, producing the observed
reflection and Fe-K fluorescence components. In the alternate,
``truncated disk'' model the thin disk is truncated at some distance
from the black hole and the inner region is filled with a hot,
radiatively-inefficient flow, which produces the hard spectrum. The
reflection spectrum and Fe-K fluorescence is then produced by the
interaction of the hard X-rays with the inner part of the truncated
disk, or in some cool outflow moving away from the disk. For a recent
discussion of the low-hard state see sect. 4 of
\cite{2007A&ARv..15....1D}.

In theory, the presence or absence of a cool disk should be
confirmable through direct detection of a soft X-ray blackbody
component at low energies. In practice however, this is made difficult
by the fact that at low accretion rates the temperature of even an
untruncated disk will drop from about 1-2keV in the high soft state to
$\sim$ 0.1-0.3keV, which puts it out of the range of most X-ray
detectors. Additionally, the effects of interstellar absorption become
very strong at around 0.1keV, so that detecting a soft excess and
accurately measuring its parameters will depend somewhat on how
accurately the interstellar absorption can be determined.

Even with these challenges, a soft excess in the low-hard state has
previously been reported in several sources. The first was Cyg X-1
\citep{1995A&A...302L...5B, 2001ApJ...547.1024D}, although its
association with an accretion disk is complicated by the fact that Cyg
X-1 is a high mass X-ray binary accreting from a wind. This question
was also the focus of two recent papers, \cite{2006ApJ...652L.113M,
2006ApJ...653..525M}, in which the authors studied long-exposure
{\it{XMM-Newton}} spectra of two different GBHCs, SWIFT J1753.5-0127
and GX 339-4, at low luminosities ($L_{\rm{X}}/L_{\rm{Edd}} \sim
0.003-0.05$). Soft excesses at similar luminosities in these two
sources have also been reported in \cite{2007MNRAS.378..182R}
(J1753.5-0127) and \cite{2008arXiv0802.3357T} (GX 339-4). Since these
two observations, there have also been observations of soft excesses
in several other sources. \cite{2007ApJ...666.1129R} made several
observations of the soft component of XTE J1817-330 with $Swift$
during the outburst decline of that source down to a luminosity of
$L_{\rm{X}}/L_{\rm{Edd}} \sim 0.001$, while a soft component in GRO
J1655-40 has been reported by both \cite{2006MNRAS.365.1203B} and
\cite{2008PASJ...60S..69T} using different telescopes.

To interpret the soft excesses in SWIFT J1753.5-0127 and GX
339-4, \cite{2006ApJ...652L.113M, 2006ApJ...653..525M} fit the data
with a several {\it XSPEC} models, trying various black-body disk
models and simple hard X-ray components (both a power law and various
Comptonization models). In GX 339-4 a broad Fe-K line was also
observed and fit with a relativistically broadened reflection
model. Using blackbody models for a standard accreting disk, the
authors found disks with maximum temperatures of {\it{kT}} $\sim$
0.2-0.4 keV, and inner radii consistent with the innermost stable
circular orbit of a black hole.

At the inferred low accretion rates in the hard state, a disk
extending to the last stable orbit would produce a soft X-ray
component with peak close to the cutoff due to interstellar
absorption. Unless an accurate independent measure of the interstellar
absorption column is available, spectral fitting procedures cannot
reliably distinguish between a thermal peak at $kT = 0.3$ keV with one
interstellar absorption column and a cooler component with a lower
energy component cutoff by a slightly higher interstellar absorption
column.

For energetic reasons the hard X-ray component which dominates 
the luminosity in the hard state must originate near 
the black hole, the same region as the proposed cool disk. Some
form of interaction of hard X-rays with the cool disk must take place,
and this implies that the isolated cool disk models used as 
`components' in fits to observed spectra are unrealistic. In fact, 
most models for the hard X-ray component include some prescription 
for the reprocessing of hard into soft radiation,
whether these be truncated disks or extended disk models. As shown
by Haardt and Maraschi (1991), such models generically produce 
a similar energy flux in soft and hard X-rays. A strong soft
component is thus a natural consequence in truncated as well as
extended disk models for the hard state. 

The main difference in a truncated disk model is that the soft flux 
originates from a larger surface area and consequently has a lower
temperature, putting its spectral peak below 0.5 keV. After interstellar
absorption the soft component has a peak around 0.5 keV that can 
be mistaken for an apparent thermal peak with the temperature
of a disk near the last stable orbit.

In this paper we examine this with a more quantitative model
for truncated disks. At the inner edge of a truncated disk the
accretion flow must change in nature from a relatively cool, thin disk
into a much hotter, vertically-extended inner flow. There will thus
necessarily be some interaction between the two, either through
radiation (e.g. \citet{1991ApJ...380L..51H}) or matter exchange
\citep{1997LNP...487...67S}, or both. Our goal is to determine
whether such a model could reproduce the soft spectral components
reported by \cite{2006ApJ...652L.113M,2006ApJ...653..525M}.
We will find that disks truncated at 15--20 Schwarzschild
radii can in fact produce soft components of the observed strength
and shape. An alternative model investigating re-condensation fron am
ADAF is considered by \cite{2008arXiv0807.3402T}.

\section{Physics of Interaction Region}\label{sec:model}

\subsection{Origin of Soft Excesses}

Determining the inner radius of an accretion disk from its spectrum
relies on (among other things) an accurate understanding of how
luminosity in the disk is produced. In the higher luminosity
``high-soft'' state, (approximately) blackbody flux from the disk
dominates the spectrum. The disk's inner radius can be inferred,
provided the distance and inclination of the source are known, from
the assumption that the radiation is produced by internal viscous
dissipation in the disk as it accretes, and an assumed `color
correction' to the blackbody spectrum.

However, when the spectrum becomes dominated by very hard X-ray and
$\gamma$-ray radiation, the radial temperature and luminosity profile
of the disk will depend on the interaction between the hard radiation
and the disk. This is because most of the accretion power is now in
the hot gas producing the hard radiation, some of which will interact
with the cool disk (as is seen in the reflection spectrum and Fe
fluorescence), and may provide a substantial source of heating. This
heating of the disk surface converts the hard radiation into a soft
component. In the simplest version of this model,
\cite{1991ApJ...380L..51H} calculated the energy balance for a hot
corona covering a cool disk and found that the flux in the soft and
hard components will be roughly equal. This predicts that a
substantial soft X-ray component is a universal feature of the
hard-state spectrum, whether the disk is truncated or not. Its
detectability depends on the sensitivity of detectors in the 0.1-1 keV
range, and the interstellar absorption column.

For models in which a cool, truncated accretion disk encircles a
very hot inner flow (which cools through inverse Compton scattering),
there will necessarily be a region of interaction near the truncation
radius, which will heat the inner edge of the disk. This will come
either from hard photons bombarding the disk as in the Haardt-Maraschi
model, or through matter interaction, with hot protons from the inner
flow directly colliding with the cool disk. In both cases previous
work has shown that a soft component is produced, but a more detailed
model is needed to predict the radiation spectrum. The first case,
examining the structure of a thin accretion disk bombarded by hard
photons has been studied by for AGN disks \cite{2001MNRAS.327...10B}
and \cite{2001ApJ...546..406N}, and more recently for GBHC disks by
\cite{2007MNRAS.381.1697R}. In this paper we focus on the second case,
in which a moderately truncated disk is embedded in a hot,
two-temperature advection-dominated accretion flow (ADAF).

\subsection{Definition of the model}

We begin with the results from a prototype model for ion bombardment
on cool disks, initially proposed by \cite{1997LNP...487...67S}, and
extended in \cite{2002A&A...387..918S}, \cite{2002A&A...387..907D} and
\cite{2005A&A...434..415D} (hereafter DS05). In this model a cool disk
is embedded in a two-temperature ADAF in which the protons are close
to their virial temperature ($\sim 20$ MeV). The protons bombard the
disk, and are stopped via Coulomb collisions with the disk's
electrons, thus transferring their energy to the disk. The energy from
the protons is sufficient to evaporate the upper layers of the disk
into a hot corona with $kT_e \sim 60-80$ keV (called the ``hot
layer'' in our nomenclature, although the temperature in this
layer is still much cooler than the virial temperature), whose
temperature is set by a balance between heating from the ions and
cooling, predominantly through inverse Compton scattering of disk
photons. \cite{2000A&A...362....1D} found that the optical depth of
the hot layer is around unity, varying only weakly with irradiating
flux and distance from the hole.

\begin{figure}
\resizebox{\hsize}{!}{\includegraphics{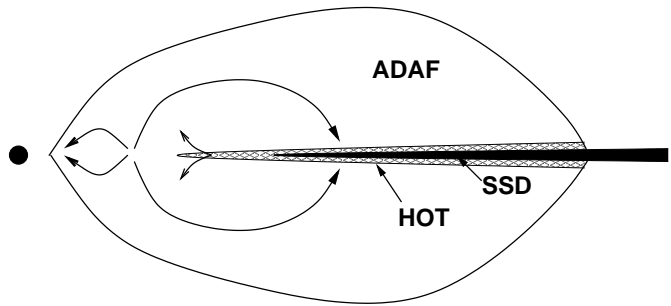}}
\caption{Schematic structure of a cool disk embedded in an ADAF. Here
the ADAF (the transparent outer region) extends over the inner edge of
the cool Shakura-Sunyaev disk (SSD), shown by the solid thin disk,
bombarding it with high energy protons ($kT_p\sim 20$ MeV). This
 evaporates the surface of the disk into a ($kT_e \sim 70$keV)
corona, called here the hot layer (HOT LAYER, hatched). The higher
viscosity of the hot layer causes it to spill over the inner edge of
the SSD where cooling through inverse Compton scattering is no longer
efficient (HOT RING), and thermal instability causes the layer to
evaporate into the ADAF. The mass transfer is represented by the small
arrows.}\label{fig:isaf} 
\end{figure}

The higher viscosity of the hot layer also allows it to spill over
inside the inner edge of the cool disk (see
fig. \ref{fig:isaf}). Using 1-D simulations, \cite{2002A&A...387..907D}
suggested that the lack of seed photons in this layer will cause the
temperature to rise to about $kT_e \sim 200-300$ keV, since cooling
from inverse Compton scattering will be much less
efficient. \cite{2002A&A...387..918S} found that this region will
become unstable and evaporate into the ADAF. This shows how an
ADAF can be maintained inside a truncated cool disk. The
key for the whole process is the presence of a component of
intermediate temperature (the hot layer). On one hand this component
produces a hard, Comptonized spectrum, while on the other its
evaporation feeds the ADAF.

DS05 extended this work to examine the radial dependence of the three
components of the flow. They set up a model in which the truncation
radius of the cool disk is a free parameter, and the three layers are
assumed to interact only via the mass transfer processes outlined
above. They assumed the cool disk was a standard Shakura-Sunyaev disk
\citep{1973A&A....24..337S}, and that the ADAF flow temperature
followed the model of \cite{1994ApJ...428L..13N}. They approximated
the temperature in the hot layer to be constant, and also assumed a
constant temperature for its extension inside the inner edge of the
cool disk (which we will call here the ``hot ring'')\footnote{The
nomenclature we adopt in this paper is slightly different from
DS05. They referred to the ADAF as an ``ISAF'', or Ion-Supported
Accretion Flow, while what we have called the ``hot inner ring'' they
term the ``hot layer'', and our ``hot surface layer'' is in that paper
referred to as the ``warm layer''.}. With these assumptions, the
radial profiles of the three components can then be determined by
solving the equations for mass and angular momentum conservation for a
thin disk.

This simplified analysis yields for each
component estimates of the mass accretion rate $\dot{M}_i$, heating
rate $Q_i$ (from internal viscous dissipation and ion heating), and
surface density $\Sigma_i$ as a function of radius. For further
discussion of the model, see sect. 2 of
DS05.\footnote{We note that in that work the
numerical factor of equation (2) is incorrect, it should be 
$2.64\times 10^8$, and equation (3) should be the same as equation (16) in
\cite{2002A&A...387..918S}.} The input parameters for the model are
the black hole mass $M_{BH}$, the magnitude of the $\alpha$ viscosity
parameter \citep{1973A&A....24..337S}, and the truncation radius
$R_{SSD}$ for the disk. A sketch of the model with its various
components is seen in fig. \ref{fig:isaf}. Assuming we know the
dominant radiation mechanisms, we can determine the spectrum from the
resulting energy distribution for each layer. This is the primary goal
of the current work.

\subsection{ Energy and mass balance}
\label{sec:endist}

The three components of the model exchange mass and energy. In the
steady state assumed here, the sources and sinks balance in each. The
topmost layer, the ADAF, loses mass and energy to the
hot layer, at rates per unit area of the order $c_{\rm s,i}\rho_ {\rm
i}$ and $c_{\rm s,i}w_ {\rm i}$, respectively. {Here $c_{\rm s,i}$,
$\rho_{\rm i}$, and $w_{\rm i}$ are respectively the thermal speed in
the ADAF (i.e. near the virial speed or orbital
velocity), the density, and enthalpy.} Depending on its radiative
efficiency the ADAF could make a substantial contribution to the hard X-ray
flux in addition to the flux from the ion-heated hot layers. For
our prototype model we ignore this contribution, since
\cite{1997ApJ...489..865E} estimate the efficiency of an ADAF to be
about $\epsilon \sim \dot{m}\alpha^2 \simeq 1.5\times 10^{-4}$ (where
$\dot{m}$ is the accretion rate scaled in terms of Eddington). If this
is the case, the luminosity of the ADAF will scale with $\dot{m}^2$,
while the energy influx into the cool disk scales as $\rho_{\rm i}
\sim \dot{m}$, (since $T \sim T_{\rm vir}$ in the ADAF), meaning that
the ADAF's indirect contribution to radiation through heating the cool
disk will dominate at low luminosities. However, the radiative
efficiency of hot inner ring is strongly model dependent, and may be
much higher, which would make the spectral contribution of the ADAF
important. We discuss this question further in sect.
\ref{sec:discussion}.

There is a mass flux by evaporation from the hot layer into the
ADAF, as derived in \cite{2002A&A...387..918S}. The
energy flux into the ADAF associated with this
evaporation can be neglected, since the temperature of the hot layer
is low compared with the temperature of the ion flow. The hot layer is
heated in similar measure by viscous dissipation and the hot
ions it absorbs from the ion flow above ($Q_{visc,\rm w} = 2\times
10^{16}\;\rm{erg s^{-1} cm^{-2}}$, and $Q_{i,\rm w} = 7\times 10^{15}
\; \rm{erg s^{-1} cm^{-2}}$, respectively for the reference model). It
is cooled by Compton upscattering of disk photons passing through it,
which will produce a hard X-ray spectrum.

There is also a mass flux feeding mass into the hot layer from
the cool disk underneath; it is parametrized as in sect. 2.2 of
DS05. The cool disk is heated in three ways: by internal viscous
dissipation, by reprocessing of hard (Comptonized) X-rays from the hot
layer above, and by the (small) number of hot ions from the
ADAF that pass through the hot layer.

The reprocessing of the hard X-rays is further divided into a fraction
$a$ that is {\em reflected}, and a fraction $(1-a)$ that is absorbed
and assumed to thermalize completely. The reflection process and the
spectrum resulting from it are based on results from the literature;
this is described in sect. \ref{sec:ref}. The spectrum of the cool
disk feeding photons to the Comptonization process in the hot layer
thus consists of a blackbody component plus a reflection spectrum.

The surface temperature of the cool disk, assumed to be close to
its effective temperture, is given by
\begin{equation}
{\sigma_B}T_{eff} ^4= Q_{\rm c, visc} +(1-a)F_{\rm h}^- +f_{\rm
c}Q_{\rm i},
\end{equation}
where $Q_{\rm c, visc}$ is the viscous heating rate of the cool disk
$_{\rm c}$, $a$ its X-ray reflection albedo, $F_{\rm h}^-$ the flux of
downward directed X-ray photons from the hot layer $_{\rm h}$, and
$f_{\rm c}Q_{\rm i}$ the fraction $f_{\rm c}$ of the energy flux
$Q_{\rm i}$ in hot ions that make it through the hot layer and are
absorbed instead in the cool disk. The albedo $a$ of the disk depends
on how ionized the disk is, which we discuss more in
sec. \ref{sec:ref}.

Compton upscattering of the soft flux from the cool disk
determines the detailed shape of the spectrum emerging from the hot
layer. In addition there is an X-ray component from the `hot ring'
interior to the inner edge of the cool disk. Since it receives fewer
of the soft photons and hence is hotter, it adds a harder component to
the spectrum.

These ingredients are the same as in DS05, except that the reflection
process is included more realistically, and the Comptonization of soft
photons is treated in detail with the Monte Carlo code of Giannios
(see sec. \ref{sec:comparison}) so a realistic X-ray spectrum is
obtained.

\subsection{Reference model}

 We illustrate the model first with representative parameter values
for a black hole binary. In sects \ref{sec:SWIFT} and
\ref{sec:GX339-4} the parameters are adjusted for application to
observations of specific objects.

We take $M_{BH} = 10 M_\odot$ for the mass, $R_{in}= 20\, R_{\rm{S}}$
for the truncation radius (where $R_{\rm S}=2GM/c^2$ is the
Schwarzschild radius). A typical if somewhat large viscosity
parameter $\alpha = 0.2$ is used for all accretion components. The
flow model of DS05 yields an accretion rate of $\dot{M} = 3.1 \times
10^{16}$ g/s for these parameter values. In terms of the Eddington
accretion rate (defined here with an assumed bolometric efficiency of
10\%) this translates to $\dot{M} = 2.2\times 10^{-3}\dot{M}_{\rm
Edd}$\footnote{In DS05 a plotting error resulted in the mass
accretion rates in all figures being overstated by a factor of 100
with respect to the Eddington rate.}, and bolometric
luminosity is $L_{\rm{X}} = 4.2\times 10^{-4} L_{\rm{Edd}}$, which is
significantly lower than observed luminosities for the low/hard
state. We discuss this discrepancy and possible solution further in
sects \ref{sec:comparison} and \ref{sec:discussion}.

\begin{figure}
\resizebox{\hsize}{!}{\includegraphics{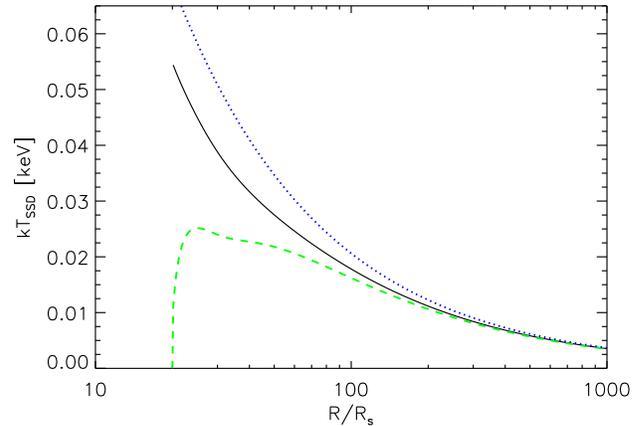}}
\caption{Effect of ion-bombardment on the surface temperature of a
cool (SSD) disk with inner edge at $20R_{\rm S}$ (for the other
model parameters see text). Black solid line: radial temperature
profile when disk heating is considered. 
Green dashed line: radial temperature profile in disk without
considering heating from hot layer.
Blue dotted line: radial temperature profile for an untruncated disk
at the same accretion rate.}
\label{fig:temp}
\end{figure}

In fig. \ref{fig:temp} we show the temperature profile of the
 cool disk component of this reference model. The dashed line shows
 the temperature profile that would result if only the viscous
 dissipation in the cool disk were included. The cool disk is
 completely evaporated into the hot layer before the inner edge of the
 accretion disk, hence its surface density drops to zero at the inner
 edge of the disk in the same way as for a standard cool disk
 accreting on a slowly rotating object (the standard ``zero-torque
 inner boundary condition''). If there were viscous coupling between the
 inner edge of the disk and the hot ring, however, the resulting
 temperatures near the inner edge would be higher than in our current
 model. This would result in a stronger soft component, whose
 contribution would remain observable for larger truncation radii than
 we have considered here. The solid line shows the effect when
 heating from the hot layer is included. As can be seen in the
 figure, the effect of heating extends to large radii. The dotted line
 shows the temperature profile for an standard untruncated disk for
 the same accretion rate.

 We can also compute the bolometric radiative efficiency for the
 model, assuming that the hot layer and inner ring radiate
 efficiently. For our reference model with a disk truncated at 20
 R$_{\rm{S}}$, we find a radiative efficiency (the ratio of the total
 luminosity to the accreting rest-mass energy flux) of about 4\%. In
 comparison, the radiative efficiency of a disk truncated at 20
 R$_{\rm{S}}$ alone is about 1\%, which demonstrates that the
 ion-bombardment process can significantly increase the radiative
 efficiency.

\begin{figure}
\resizebox{\hsize}{!}{\includegraphics{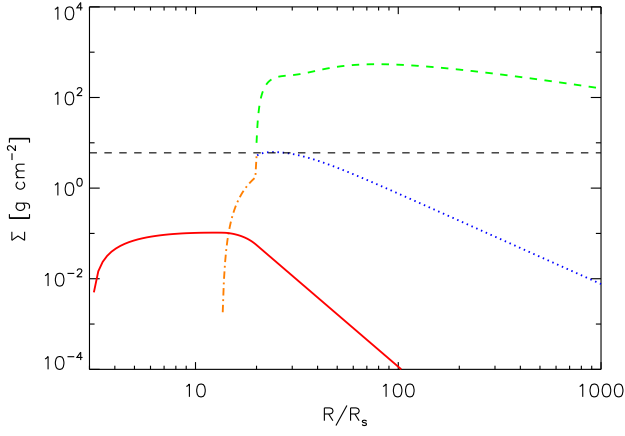}}
\caption{Surface density ($\Sigma$) profile for reference model, where
$M_{BH}= 10 M_\odot$, $\alpha = 0.2$ and $R_{ssd} = 20 R_s$. Green
dashed line: Shakura-Sunyaev disk. Blue dotted line: Hot
layer. Orange dash-dotted line: Hot ring. Red solid line:
ADAF.}\label{fig:sigma}
\end{figure}

\subsubsection{Cool disk emission}
\label{sec:cooldisk}

We begin by estimating the spectrum of the radiation produced by the
thin disk, i.e. the seed photons of the Comptonization in the hot
layer. At large effective optical depths in the disk the radiation
field is a blackbody. The flux emerging from the photosphere of the
disk is reduced, however, by Thomson scattering, resulting in a `modified blackbody', a spectrum with same
temperature but reduced flux. In the literature, it is often described
equivalently by a `color correction' factor $f_{\rm col}$, measuring
the effect instead as the ratio of color temperature to effective
temperature, $T_{\rm col}=f_{\rm col}T_{\rm eff}$.
Several papers have done extensive calculations of this spectral
hardening in the outer layers of a disk dominated by electron
scattering, including \cite{1995ApJ...445..780S},
\cite{2000MNRAS.313..193M}, and \cite{2005ApJ...621..372D}, and
concluded that the the color-correction is of the order $f_{\rm col}
\simeq 1.4-2$, with the magnitude of correction increasing gradually
as the luminosity increases. The approximation of a single parameter
across the entire spectrum become less good as the flux decreases.

In a cool disk the (vertically integrated) viscous dissipation 
$Q_{\rm visc}$ is balanced by radiative loss at its surface, 

\begin{equation}
Q_{\rm visc}=\sigma_B T_{\rm eff}^4= \sigma_B(T_{\rm col}/f_{\rm
col})^4.\label{fcol}
\end{equation}

The value for $f_{\rm col}$ will change the determination of the inner
disk radius, since the luminosity will be depressed by a factor of
$f_{\rm col}^{-4}$ relative to the temperature, so that when color
correction is considered, the inner radius of the accretion disk will
be increased by a factor $f_{\rm col}^2$.

In this work we have the additional complication of the hot surface
layer, which will heat the upper layers of the disk substantially and
may change the spectral hardening. \cite{2007MNRAS.381.1697R} have
estimated the effects of incident hard X-ray radiation on galactic
black hole disks, although they focus on systems in which the disk is
hotter ($kT \sim 0.35$ keV) than ours and dominates the overall
spectrum. However, they do find substantial spectral hardening, and
disk temperatures which increase with increasing X-ray flux
(fig. 4 of that paper). They also find that the incident radiation
can change the ionization structure of the upper layers of the cool
disk, so that the shape of the disk spectrum is significantly altered
by absorption lines (see their fig. 6).

Since it is beyond the scope of this work to do a similar radiation
transfer calculation, we instead choose a single color correction
factor, $f_{col} = 1.7$ as a typical value for all our models. 

\subsubsection{Reflection spectrum}
\label{sec:ref}

The hard flux incident on the cool disk undergoes reprocessing by
electron scattering and atomic processes, giving rise to a series of
emission lines (especially the Fe-K fluorescence line) and a Compton
reflection hump. In the course of these processes, part of the
incident energy flux ends up heating the plasma. Detailed
calculations of these processes is beyond the scope here. Instead, we
simplify the physics by separating the incident flux into a reflected
part and a part that is treated as being absorbed. We treat the
absorbed part in the same way as the energy input by viscous
dissipation in the disk, i.e. a fraction $(1-a)$ is added to the left
hand side in (\ref{fcol}). 

The remaining part is treated by a modified version of the XSPEC
model ``REFLION'' described in \citep{2005MNRAS.358..211R}. This
model calculates the reflection spectrum of a power law flux between
0.1-300 keV hitting a constant density AGN disk. We set the lower
cutoff of this power law at 0.2 keV, since we expect disk
temperatures around $0.1-0.2$ keV. REFLION calculates the
reflection spectrum for a fixed ionization parameter, $\xi \equiv 4\pi
F_{\rm{X}} n_H^{-1}$, the ratio of the incident flux on the disk to
the hydrogen number density. The disk, however, is stratified, so an
estimate has to be made of the depth in the disk atmosphere where most
of the reprocessing takes place. We assume here that this depth is
just the electron scattering photosphere of the cool disk, $\tau_{\rm
es}=1$. By hydrostatic balance, the pressure at this depth is
approximately

\begin{equation}
P \simeq \frac{g}{\kappa_{es}},
\end{equation}

where $g$ is the vertical component of the acceleration of gravity at
the height of $\tau_{\rm es}=1$ above the midplane. Assuming this
height is about twice the nominal disk thickness $H=c_{\rm
s}/\Omega_{\rm K}$, the pressure is $P\simeq 2c_{\rm s}\Omega_{\rm
K}$. Here $c_{\rm s}$ is the sound speed at the
midplane of the disk, which is determined from the heating rate
$Q_{\rm visc}$ by a standard thin disk model. To fix the density
corresponding to this pressure the temperature has to be
determined. We assume for this the effective surface temperature
$T_{\rm eff}$ that follows from the viscous heating rate. If
 $F_{\rm h}^-$ is the incident X-ray flux from the hot layer, the
ionization parameter is then

\begin{equation}
\xi \simeq \frac{2\pi F_{\rm h}^- kT_{eff}\kappa_{es}}{\Omega_K c_s}.
\end{equation}

 The value of $\xi$ varies with distance $r$ from the black hole; an
energy-weighted mean is $\xi\simeq 10$ erg cm s$^{-1}$ for the
reference model. We use this as a representative value for the
reflection calculation instead of an integration over the disk. For
this value of $\xi$, the resulting albedo is low ($\simeq$ 0.2), so
that most of the hard flux is absorbed and thermalized in the
disk. The total reflection and disk spectrum (before being passed
through the hot layer), is seen in the red long-dashed line in
fig. \ref{fig:TESTspec}. This is the input spectrum for the
Comptonization by the hot layer. Since the optical depth of the hot
layer is only around unity, it will also make a significant direct
contribution to the output spectrum.

\subsubsection{Uncertainties in the reflection spectrum}

 Through its effect on the gas density, the temperature assumed in
the reprocessing region has a direct effect on the ionization
parameter and the resulting reflection spectrum. Unfortunately, there
is a significant uncertainty in this temperature, since it depends
itself on the ionization parameter. At low densities higher in the
atmosphere, the ionization parameter is high and the temperature will
be close to the Compton temperature of the incident X-ray spectrum.
For our reference model, this would be around 3 keV. Deeper in the
atmosphere, the ionization parameter would be lower and the
temperature closer to equilibrium with the energy density of the
radiation field, of the order of the effective temperature of the cool
disk. Since the transition between these regimes takes place around
the reprocessing layer itself, all depends on details of the radiation
physics.

Several groups \citep{2001MNRAS.327...10B,2000ApJ...537..833N} have
calculated the change in the vertical structure of AGN disks as a
result of incident hard radiation, and found the upper layers become
stratified, with the top being heated to close to the Compton
temperature and a sharp transition to the inner layer which radiates
at the disk's effective temperature. They find that this
stratification will change the disk's reflection properties
substantially. Although \cite{2007MNRAS.381.1697R} study a similar
situation in galactic black holes, and find no similar stratification,
they focus on cases in which the disk dominates the luminosity. For
lower disk temperatures stratification could presumably still
occur. However, \cite{2001ApJ...546..406N} studied the case where a
disk was overlaid with a surface corona, and included the effect of
its weight on the gas pressure. They found that this was sufficient to
prevent stratification.

An additional caveat in using the constant density models of
\cite{2005MNRAS.358..211R} is that they consider AGN disks, which are 
much cooler and less dense than those in GBHCs. As a result, the reflection 
spectra do not consider the flux in the disk itself (which can change the 
ionization state significantly), and the effects from things like three-body 
recombination, which will become important as the disk densities become 
higher. Recent work by \cite{2007MNRAS.381.1697R} has examined
reflection spectra in GBHCs, although they focus on systems in
which the disk dominates the spectrum and is hotter ($kT_e = 0.35$ keV)
than the disks considered in this work. 

\subsubsection{Comptonization}
\label{sec:comp}

To calculate the spectrum from inverse Compton scattering through the
hot layers of the disk we use a one-dimensional Monte Carlo
simulation. The radial variation of quantities in the disk and hot layer 
is replaced by representative values, for which we use energy-weighted 
means. The simulation assumes a slab geometry, with the cool
optically thick disk below a much hotter surface layer with moderate
optical depth. For the seed photons, we use the disk+reflection spectrum 
found in the previous two sections. To capture the emission lines in
this spectrum, we set the resolution of the simulation to 
$\Delta E/E = 0.046$. The number of seed photons followed through 
the hot layer is of the order $10^7$, sufficient to represent the
result to a noise level comparable with typical observations. The
output spectrum is angle dependent (because of the increasing optical
depth with inclination), so a value has to be assumed for the inclination 
angle $\theta$ to the line of sight (measured from the normal of the disk). 
For our reference model we use $\mu=\cos\theta = 0.5$ as a 
representative value. 

 The input parameters of the Comptonization calculation are the input
spectrum, the optical depth and the temperature of the hot layer. The
resulting total energy flux in Comptonized photons (integrated over
the spectrum and summed over both sides of the hot layer) has to match
the energy input into the hot layer: the sum of ion heating and
viscous dissipation. We can estimate the optical depth for the
hot layer from the surface density of the hot layer, where
$\langle\tau\rangle = \kappa_{es}\langle\Sigma_{\rm{Hot}}\rangle$.
The temperature of the hot layer is adjusted iteratively in the
calculations to meet this condition to an accuracy of 10\%. For the
reference model, we find $kT_e = 70$keV, and $\tau = 0.87$. The
resulting emergent spectrum of the hot layer is shown in the blue
dotted line in fig. \ref{fig:TESTspec}, for an inclination angle of
$\mu = 0.5$. The photon index of this spectrum in the range 2-10 keV
is about $\Gamma = 1.96$.

\subsubsection{Treatment of the hot ring}

In the hot ring (where the hot layer has spilled over inside inner
edge of the cool disk) there is no underlying cool disk any more and it
receives its soft photons only by scattering from larger distances: it
is `photon starved' compared with the hot layer itself. Since the
energy input by ion heating and viscous dissipation are still similar,
the temperature is higher and Comptonization correspondingly
stronger. The hot ring therefore makes its contribution mostly at the
high energy end of the spectrum, and is less important for the `soft
component' of the spectrum that is the focus of our study. We include
it, however, since we also want to achieve a reasonable fit to the
overall spectral energy distribution in the observations discussed in
the next sections.

The calculation of DS05 relied on the earlier one-dimensional work of
\cite{2002A&A...387..907D} in order to set the temperature and
energetic contribution from the hot inner ring, which predicts a very
small contribution from the hot ring. With a sufficiently detailed
geometrical model for the hot ring, the soft photon input by
scattering could in principle be modelled more accurately, but the
level of detail needed is probably beyond the limits of the present
model. We therefore treat the soft input flux in this component of
the model as an adjustable unknown when fitting to spectra. For this,
we introduce a parameter $\zeta$, which represents the fraction of
seed photons from the cool disk that cool the hot layer, so that
$1-\zeta$ goes to cool the hot ring. For the reference model, $\zeta$
is set by the contribution predicted from the DS05 model for the hot
inner ring.

Also treated as adjustable is the temperature reached by equilibrium
between heating and Comptonization. We assume that cooling in this
region is still moderately efficient, choosing temperatures in the
range $kT_{\rm{e}}= 180-200$ keV. For simplicity (since the angular
distribution of seed photons is also uncertain) we model the hot
ring with a plane-parallel Monte Carlo simulation with the same
resolution as in the hot layer. We discuss the limitations of this
approach in sect. \ref{sec:discussion}.

 As in the hot layer, the X-ray energy flux produced by
Comptonization has to match the energy input by viscous dissipation
and ion heating. Together with the now assumed value for the
temperature, this determines $\zeta$, and the optical depth of the
ring. For the reference parameters of this section, we find
$\zeta=0.99$ and an optical depth of $\sim 0.7$. At a temperature of
200 keV we find that the ring contributes only 11\% of the
overall hard flux (F$_{\rm X}= 2-200$ keV) for the reference values
$\alpha=0.2$ of the viscosity and $R_{in}=20R_S$.

This is because the radial width of the hot ring has a rather limited
extent, since the evaporation process is very efficient. The hot layer
evaporates very quickly after it has flowed over inner edge of the
cool disk. The hot Comptonized component from the hot ring is shown in
the orange dash-dotted line in fig. \ref{fig:TESTspec}.

\subsection{Results for the reference model}
\begin{figure}
\resizebox{\hsize}{!}{\includegraphics{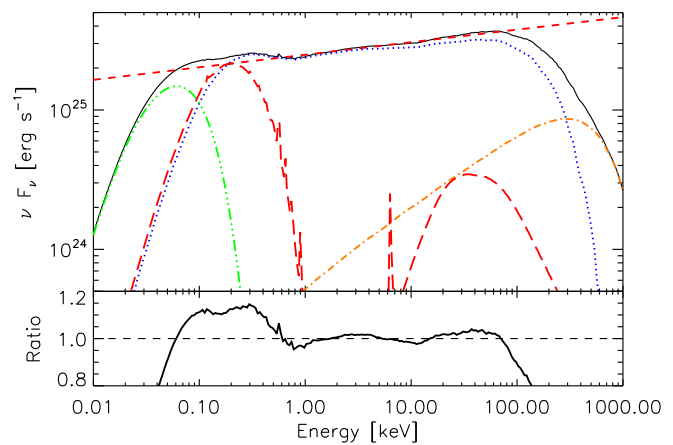}}
\caption{Top: Relative contibution from each component for
reference model, $\alpha = 0.2$, $R_{in} = 20R_s$, $M=10M_\odot$. Red
long-dashed line: Spectrum from the disk (modified blackbody plus
reflection spectrum). Green dash-double dotted line: Spectrum from
outer disk. Blue dotted line: Comptonized spectrum from hot
layer. Orange dash-dotted line: Comptonized spectrum from hot
ring. Black: total spectrum. Red short-dashed: power law with $\Gamma
= 1.9$. Bottom: Total spectrum divided by power law with $\Gamma =
1.9$.}\label{fig:TESTspec}
\end{figure}

The final spectrum and its various components is shown in the top
panel of fig \ref{fig:TESTspec}. The accretion rate for the
reference model is $\dot{M}/\dot{M}_{\rm{Edd}} = 0.002$ assuming an
efficiency of 10\%, while the luminosity in the 0.5-10 keV range is
$L_{\rm{X}}/L_{\rm{Edd}} = 10^{-4}$. The total spectrum is shown in
black. The individual components run as follows. The green thermal
component (dash-double dotted line) shows the spectral contribution
from the outer part of the disk where the hot layer is no longer
significant (outside $R/R_S = 100$), while the red long-dashed
component shows the rest of the disk and reflection spectrum. For the
reference model the reflection ionization parameter is small enough
that the reflection and iron line are not apparent in the final
spectrum, although we again stress that we are using a reflection
model developed for AGN, so in reality the reflection could be
stronger. The blue dotted line shows the Comptonized spectrum from the
hot layer, while the orange dash-dotted line shows the Comptonized
spectrum from the hot ring. Fitting the overall spectrum with a
photon index $\Gamma = 1.91$ in the 1-10 keV range, we see a small
soft excess below 0.5keV, even though the maximum disk
temperature is only 0.05keV. The bottom panel of fig
\ref{fig:TESTspec} shows the total spectrum divided by a power law
with $\Gamma = 1.91$. The observed deviation from a single power law
in the hard part of the spectrum (which leads to a deficit
around 1 keV and a harder power law index above 10 keV) is caused by
anisotropic Comptonization resulting from considering a plane-parallel
configuration (see e.g. \cite{1993ApJ...413..680H} and sect.
\ref{sec:discussion} of this paper), and also by the
contribution from the hot ring. Except for the very low
accretion rate, we see a spectrum that is qualitatively similar to
those of \cite{2006ApJ...652L.113M} and \cite{2006ApJ...653..525M}.

\section{Comparison to observations}
\label{sec:comparison}

The luminosity of our reference model described above, of the order
$10^{-4}L_{\rm Edd}$ is substantially lower than the luminosities (of
the order $3\times 10^{-2}$-$10^{-3}L_{\rm Edd}$) inferred for the
observed sources to which we want to apply the model. This is a
consequence of the flow model in DS05 that is the basis of our
analysis. In it, the surface density of the hot layer is governed by
the physics of the Coulomb interaction of the hot ions penetrating
through it, and its temperature by the energy balance between it and
the underlying disk. With temperature and surface density constrained
in this way, the mass flux then depends only on the radial drift
speed, i.e. the viscosity parameter, $\alpha$. The actual mass flux is
low because the temperature of the layer is only about 80keV. This
suggests that the current model is incomplete, and we discuss a
possible solution in sect. \ref{sec:discussion}. In this paper,
however, we solve this problem by introducing a parameter, $C$, by
which the accretion rate (or equivalently the energy output of each
component of the flow) is increased. This makes the implicit
assumption that increasing the accretion rate causes the energy output
in each component to increase, but the relative contribution of each
component to the overall energy budget  and geometric
configuration of the flow to stay constant. Scaling the
accretion rate in this way increases the luminosity in all
components, which thus causes an increase in temperature
for the cool disk at a fixed truncation radius.

With the introduction of the parameters $C$ and $\zeta$
our model loses its predictive power, but the goal of this paper is to
present a model that is plausible rather than precise in its
details. In the next section we compare our spectra to the best fits
from observations, and show that for reasonable values of the hot
layer and inner ring and accretion rate, we can reproduce the observed
soft excesses using a significantly truncated accretion disk.

To illustrate how our model can be made consistent with observed
soft excesses, we perform a qualitative comparison between the soft
excesses observed in SWIFT J1753.5-0127 \citep{2006ApJ...652L.113M}
and GX 339-4 \citep{2006ApJ...653..525M} and our model. For each
object we use the estimates for black hole mass, inclination and
distance to source presented in those two papers and take $\alpha =
0.2$ as a standard value for the viscosity parameter. We then assume a
 moderate truncation radius and find a solution for energy and
surface density as was discussed in sect. \ref{sec:model}, and
calculate the spectrum. We change $C$ (the accretion rate) and $\zeta$
(the ratio between seed photons in the hot layer and inner ring) in order
to match the luminosity and spectral index of the best fit to the
observed spectrum, and compare our soft excess to the observed one. If
necessary we also change the amount of interstellar absorption,
although for both the cases we consider we do not need to change it
very much. Given the systematic uncertainties in our model, a
more statistical comparison to the data is not possible; our goal is
instead to demonstrate that we are able to reproduce the observed
spectra with physically reasonably parameters.
 
\subsection{SWIFT J1753.5-0127}
\label{sec:SWIFT}
We begin with the spectrum from the source SWIFT
J1753.5-0127. \cite{2006ApJ...652L.113M} took a 42 ksec
{\it XMM-Newton} observation and estimated an X-ray luminosity
(0.5-10 keV) of $L_{\rm{X}}/L_{\rm{Edd}} = 2.6 \times 10^{-3} (d/8.5\,
$kpc$)^2 ($M/10 M$_\odot)$. They fit the spectrum to a power law with
a photon index $\Gamma = 1.67$ (2-10 keV), and interstellar absorption
of $N_{\rm{H}} = 2.3 \times 10^{21}$ cm$^{-2}$. Fitting the spectrum
with an absorped power law component alone reveals a small soft excess
below 2 keV, which they fit to a disk with $kT_{in} \simeq 0.22$ keV
and R$_{in}\simeq$ R$_{\rm{S}}$ (M/10M$_\odot)\;
(d/8.5$kpc$)/\cos^{1/2}i$.

A very truncated disk will be too cool to be observable in
X-rays, while an untruncated disk will have a higher temperature than
is observed (because of the effects of heating from the corona). To
compare with the observed spectra we assume a moderate truncation
radius of 15R$_{\rm S}$, which is qualitatively very different from
an untruncated disk (which will have an inner radius between
$0.5-3$R$_{\rm S}$ depending on the spin of the black hole). We make
the same assumptions for mass (M = 10M$_\odot$) and distance ($d =
8.5$kpc) as in \cite{2006ApJ...652L.113M} and assume an inclination of
$\mu = 0.5$. The simulation then predicts an accretion rate of
$\dot{M} / \dot{M}_{\rm Edd } = 1.5 \times 10^{-3}$ for an efficiency
of 10\%. In order to match the measured flux in the power law
component, we increase the flux in each component by a factor $C =
12$, to give an accretion rate $\dot{M} / \dot{M}_{\rm{Edd}} =
1.7 \times 10^{-2}$.

In the hot layer, the predicted surface density profile gives the
optical depth $\langle\tau\rangle = 0.87$, and the energy balance
between the cool disk and hot layer (see sect. \ref{sec:endist})
determines a temperature of $kT_e = $ 75 keV, which gives a photon index of
$\Gamma = 1.87$ (3-10 keV).\footnote{The spectral index of a
Comptonizing corona will change depending on viewing angle, since the
mean optical depth will change depending on whether the disk is
face-on or tilted, which will harden the spectrum. To calculate the
energy in the hot layer and thus its temperature we take the spectral
index for the spectrum integrated from $0^\circ$-$180^\circ$.} The
density of the disk and flux in the hot layer also allows us to
estimate the ionization parameter, $\xi = 50$ erg cm s$^{-1}$, from
which we get an albedo and find that the Comptonized flux incident on
the disk heats it to $kT_{e} = 0.11$ keV, which looks like $kT_{e} =
0.19$ keV when spectral hardening is taken into account.

\begin{figure}
\resizebox{\hsize}{!}{\includegraphics{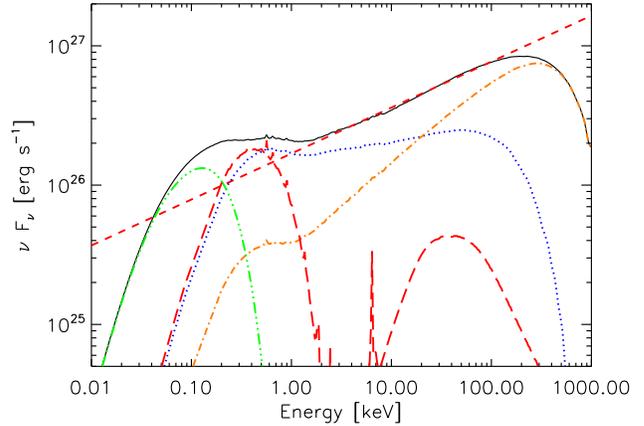}}
\caption{Model spectrum for SWIFT J1753.5-0127, for a truncated
disk with inner radius of 15 R$_{\rm{S}}$ with $\alpha = 0.2$ and
$M=10M_\odot$. Red long-dashed line: Spectrum from the disk (modified
blackbody plus reflection spectrum). Green dash-double dotted line:
Spectrum from outer disk. Blue dotted line: Comptonized spectrum from
hot layer. Orange dash-dotted line: Comptonized spectrum from hot
ring. Black: total spectrum. Red short-dashed: power law with $\Gamma
= 1.66$.}\label{fig:SWIFTspec}
\end{figure}

The predicted surface density in the hot inner ring also allows us to
calculate the optical depth of this layer, which we find to be
$\langle\tau\rangle = 0.7$. Since we do not have a detailed model for
the radiative transfer in this component, we take a plausible value of
$kT = 200$ keV, which gives a photon index of $\Gamma = 1.39$ between
2 and 10 keV. The resulting spectrum is still too soft, so we set
$\zeta =0.89$, meaning that 89\% of the disk and reflection photons
are used to seed the hot layer, while the rest seed the hot inner
ring. Figure \ref{fig:SWIFTspec} shows the resulting
spectral energy distribution for SWIFT J1753.5-0127. The different
components shown in the figure are the same as in fig.
\ref{fig:TESTspec}.

\begin{figure}
\resizebox{\hsize}{!}{\includegraphics{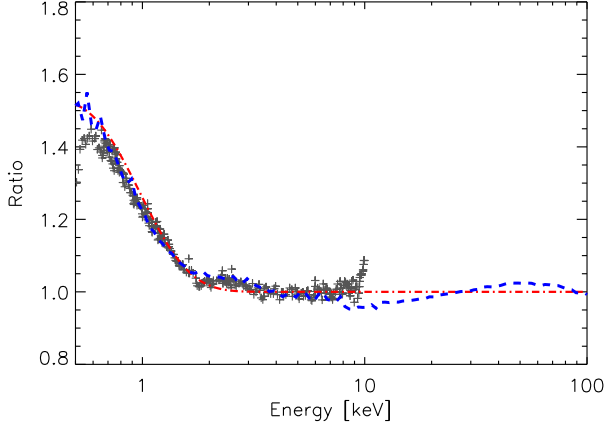}}
\caption{The soft excess in SWIFT J1753.5-0127. The crosses show the
observed spectrum divided by a simple power law fit above 3 keV with a
best-fit column density of $N_{\rm{H}} = 2.3 \times 10^{21}$cm$^{-2}$,
taken from fig. 2 of \cite{2006ApJ...652L.113M}. The
dashed-dotted red line shows the best fit two-component XSPEC
``diskbb+pow'' divided by the same power law (as reported in
\cite{2006ApJ...652L.113M}). The dashed blue line shows the
ratio between our best fit model (see fig. \ref{fig:SWIFTspec}) and
the same power law, with an increase of 0.05$\times 10^{21}$cm$^{-2}$
in interstellar absorption.}\label{fig:SWIFTcomp}
\end{figure}

Figure \ref{fig:SWIFTcomp} (taken from \cite{2006ApJ...652L.113M})
shows the spectrum divided by an absorped power law with an absorption
column density of $N_{\rm{H}} = 2.3\times 10^{21} $cm$^{-2}.$ Overlaid
we show their best fit using the XSPEC ``diskbb+pow'' model (in red)
and the soft excess predicted by our model (in green). To obtain a
better fit we change the absorption column density to $N_{\rm{H}} =
2.35\times 10^{21}$cm$^{-2}$ to render the two excesses
effectively indistinguishable, even though the temperature and peak
fluxes of both disks are very different. We discuss the reasons for
this at the beginning of sect. \ref{sec:otherwork}. Figure
\ref{fig:SWIFTcomp} also shows that the deviation from a simple
power law in the range 2-100 keV is less than 10\%, which is consistent
with observation.

\subsection{GX 339-4}
\label{sec:GX339-4}
Our second source for comparison is the relatively better constrained
X-ray source GX 339-4. This source has an estimated mass of
6$M_\odot$, distance of around 8 kpc and inclination $\mu =
0.9$. \cite{2006ApJ...653..525M} took a 280ks {\it{XMM-Newton}}
observation of this source, which they observed at a flux of about
$L_{\rm{X}}/L_{\rm{Edd}} \simeq 0.05
($M$/10$M$_\odot)(D/8$kpc$)$. They fit the data with a moderate
absorption column density ($N_{\rm{H}} = 3.72\times 10^{21}$cm$^{-2}$)
and very hard power law ($\Gamma = 1.44$) and find a soft excess below
3 keV which they fit to a disk with R$_{in} = 0.6$R$_{\rm{S}}$ and
$kT_{in} = 0.38$ keV. Additionally, they observe a broad asymmetric
Fe-K line with a maximum at 6.9 keV, from which they measure a
reflection fraction of about 0.2-0.3 and ionization parameter $\xi
\sim 10^3 $ erg cm s$^{-1}$. Fitting the Fe-K line using relativistic
broadening suggests an inner radius of 0.7R$_{\rm{S}}$.

\begin{figure}
\resizebox{\hsize}{!}{\includegraphics{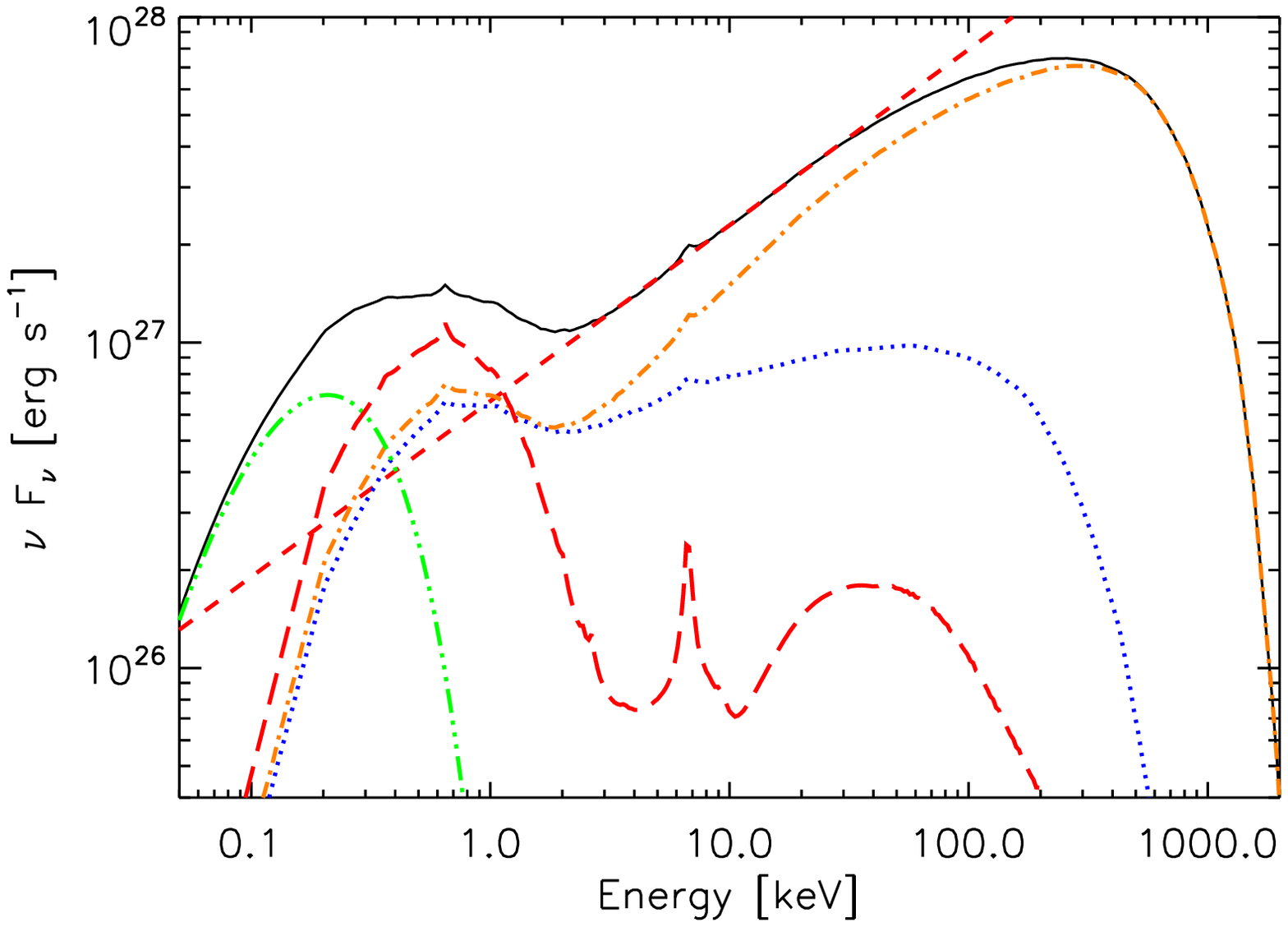}}
\caption{Model spectrum for GX339-4, for a truncated disk with an
inner radius of 19$R_S$, $\alpha = 0.2$ and $M=6 M_\odot$. Red
long-dashed line: Spectrum from the disk (modified blackbody plus
reflection spectrum). Green dash-double dotted line: Spectrum from
outer disk. Blue dotted line: Comptonized spectrum from hot
layer. Orange dash-dotted line: Comptonized spectrum from hot
ring. Black: total spectrum. Red short-dashed: power law with $\Gamma
= 1.47$.}\label{fig:GX3394spec}
\end{figure}

\begin{figure}
\resizebox{\hsize}{!}{\includegraphics{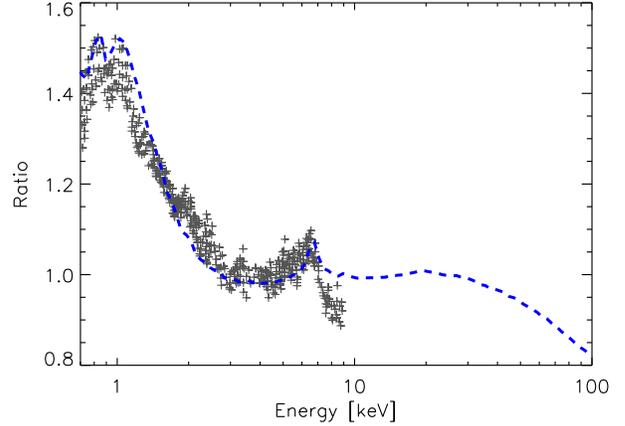}}
\caption{The soft excess in GX 339-4, taken from fig. 3 of
\cite{2006ApJ...653..525M}. The crosses show the spectrum divided by a
simple power law fit to the {\it{XMM-Newton}} and RXTE spectra with an
absorption column density of $N_{\rm{H}} = 3.72\times 10^{21}$
cm$^{-2}$. The 4-7 keV range, where a broad Fe-K line is clearly
visible, was omitted from the fit. The dashed blue line shows the
ratio between our best fit model (see fig. \ref{fig:GX3394spec}) and
the same power law fit with interstellar absorption $N_{\rm{H}} = 4.9
\times 10^{21}$ cm$^{-2}$.}\label{fig:GX3394comp}
\end{figure}

For an assumed radius of the inner edge of the disk $R_{\rm in}
=19R_{\rm S}$, a good fit for this source is obtained with the
following set of parameter values: C=77 (or $\dot{M}/\dot{M}_{\rm Edd}
= 0.15$), $\zeta$=0.63, and the temperature of the hot layer and hot
ring 75 keV, and 200 keV respectively. The resulting spectral energy
distribution is shown in fig. \ref{fig:GX3394spec}. Figure
\ref{fig:GX3394comp} shows the soft excess observed in
\cite{2006ApJ...653..525M}. The luminosity enhancement factor is
higher than for the SWIFT source because of the higher luminosity
inferred from the observations. The ionization parameter in the
reflection region is also higher, with $\xi=550$ erg cm s$^{-1}$.

Note the structure in the data around 1 keV, which is suggestive of
the structure of the interstellar absorption in this region of the
spectrum. In fact, we can best reproduce the observation for a column
$N_{\rm{H}} = 4.9 \times 10^{21}$cm$^{-2}$, some 30\% higher than
found by \cite{2006ApJ...653..525M}. 

As is seen in fig. \ref{fig:GX3394comp}, we also find an Fe-K
emission line in our fit, with a strength comparable to the
observations. The combination of parameters that fits the overall
spectral shape in our model therefore also fits the reflection
component of the spectrum. Since this component depends on details of
the interaction between the hot layer and the cool disk under it, this
adds some confidence in this part of the physics of our model.

\section{Discussion}\label{sec:discussion}

In the paper thus far we have produced spectra of an ion-bombardment
model with a truncated disk for Low-Hard state accretion,
incorporating a spectrally-hardened blackbody component, physical
reflection models from the literature and a Monte Carlo Comptonization
calculation. 

Detailed fits to observations have shown that the model can reproduce
the observed soft excess, high-energy spectral index and the
approximate strength of the Fe-K line in GX 339-4. We have also shown
that the structure observed in the soft excess in GX 339-4 is
consistent with interstellar absorption features, suggesting that the
absorption column density for this source might be underestimated.

However, we have also found that further work is necessary to develop
the flow model outlined in \cite{2002A&A...387..907D} and
DS05. Most significantly, a large viscosity in
the hot regions of the flow (the hot layer and ADAF) is required to
increase the accretion rate sufficiently to match observations. This
may suggest the presence of strong ordered magnetic fields in the
inner regions of the accretion flow, which are also believed to be
associated with observed jets. 

\subsection{Mass flux in the hot layer}\label{sec:uncertainties}

As was discussed in sect. \ref{sec:comparison}, the global accretion
rate in the present model is limited by the rate at which the hot
layer can flow over the cool disk. The surface density and temperature
of the hot layer are in turn narrowly constrained by the physics of
the Coulomb interaction which allow the layer to form and the energy
balance between it and the underlying disk. There are thus two ways to
increase the flow rate in the hot layer: by increasing the effective
viscosity or providing another mechanism besides viscous dissipation
to transport angular momentum.

The interaction between the ion supported ADAF and the hot layer
provides such a mechanism. The ion supported flow is partially
suppoted against gravity by gas pressure and rotates slower than
Keplerian. The mass condensing from the ADAF on the hot layer
thus acts as a sink of angular momentum, which increases the mass
flux in the hot layer. This effect was not included in DS05 and
the calculations above. An estimate of its importance can be made
by evaluating the angular momentum exchange {{\it{a posteriori}} from the
solutions in sect. 3. We find that, for the viscosity parameter
$\alpha=0.2$ assumed for the hot layer, the effect increases the
mass flux by a factor 2--3. The effect is thus significant, but not
sufficient to increase the mass flux by the factors indicated by
the comparison with observations in sect. 3.

The missing ingredient most likely to lead to the higher mass fluxes
inferred from the observations may well be a strong magnetic
field. Strong ordered magnetic fields in the inner regions of the flow
are implied by the presence of jets, especially in the hard X-ray
states discussed here. A bundle of strong ordered magnetic field held
together by a disk \citep{1974Ap&SS..28...45B} can have field
strengths well above those produced by magnetorotational
turbulence. The angular momentum exchange by interaction of such a
bundle with the disk \citep{2001MNRAS.323..587S, 2003ApJ...592.1042I,
2003PASJ...55L..69N,2005ApJ...620..878D} can be much more effective
than turbulence parametrized with a viscosity parameter $\alpha\sim
1$. This aspect is beyond the present study, and is a promising field for
further study.

\subsection{Spectrum of the hot ring}
The spectrum from the hot ring in our model is hard to predict without
a more detailed model. The uncertainty lies chiefly in the
distribution and number of seed photons available for cooling. The
more photon-starved the hot ring is, the higher its temperature will
be, and (since the evaporation rate into an ADAF scales with
$T_{\rm{e}}^2$) the smaller its contribution to the overall
spectrum. The temperature we assumed for the hot ring in practice
could be much lower, which would bring it more in line with
observations of the high energy cutoff observed in some spectra (which
suggest a maximum temperature of about $kT_{\rm{e}} \sim 150$ keV).
Additionally, the geometric distribution of seed photons will change
the structure of the Compton spectrum. This is because photons that
scatter once preferentially scatter back in the direction they were
originally travelling, and there is a deficit of photons in the first
scattering hump in the spectrum. This effect is most pronounced in the
plane-parallel case (e.g. \cite{1991ApJ...380L..51H}), but there will
also be some anisotropy in the hot ring's spectrum if the seed photons
are primarily from the disk. We have considered only seed photons from
the disk, but there may also be photons produced from other processes
(such as synchrotron emission) which would allow the hot ring to cool
more efficiently and make the effects of anisotropy less pronounced
(since the seed photons would be travelling through the hot ring in
essentially random directions). Finally, in this paper we have
neglected the spectral contribution from the ADAF. Depending on its
radiative efficiency, its contribution could also harden the observed
high-energy Comptonized spectrum considerably.

\subsection{Comparison with other work}\label{sec:otherwork}

Our model of a disk truncated at 15-20 $R_{\rm S}$ and surrounding
corona is qualitatively very different from the untruncated disk (with
$R_{in} \sim 1 R_{\rm S}$) models fit by \cite{2006ApJ...652L.113M} and
\cite{2006ApJ...653..525M}, and it is natural to ask how the observed soft
excess can be so small when the radiating area is so much larger. The
answer lies in several points. The most important of these is that the
temperature in our disks is about a factor 2 smaller than is found by
\cite{2006ApJ...652L.113M}, so that the flux is intrinsically much
smaller and (even after the colour-correction is applied) most of the
flux is cut off by interstellar absorption. There is a further
reduction from the hot surface layer, which upscatters about
two-thirds of the photons. Finally, the shape of the upscattered
photons deviates from a power law at low energies, so that measuring
the temperature of the soft excess depends very sensitively on
modelling the Comptonized spectrum correctly.

The effects of irradiation on the measured truncation radius have
also been studied using a more phenomenological approach in
\cite{2008MNRAS.tmp..717G}, who re-analyzed the data from J1817-330
\citep{2007ApJ...666.1129R} to demonstrate that irradiation can
increase the measured truncation radius in this source (although they
assume continual stress at the inner boundary of the truncated disk,
which we have not done here). In particular, they note future plans to
test the effects of incomplete thermalization from incident radiation
(cf. sect. \ref{sec:cooldisk}), which can further increase the
truncation radius.

Several well-studied sources show some evidence of deviations from a
single power law. \cite{2007A&ARv..15....1D} notes that spectra from
Cyg X-1 have additional structure in their spectra that can be fit
with an additional very soft Comptonizing component, while both the
source GX 339-4 and Cyg X-1 sometimes show an excess of very high
energy photons compared to a fit with a single power law, which
suggests a second site for Comptonization that is naturally explained
with this model.

For GX 339-4, \cite{2006ApJ...653..525M} detected a broad Fe-K line,
which they fitted with a relativistically broadened profile, implying
an untruncated disk and a spinning black hole. However, the
observation of a broadened Fe-K line may also be consistent with a
truncated disk if the broadening instead comes from an outflow, as has
been suggested in \cite{2006MNRAS.367..659D} and
\cite{2007ApJ...656.1056L}.

The truncated disk picture of hard states in X-ray binaries has
received support from analyses of the noise spectrum of the X-ray
variability. In the model by \cite{2001MNRAS.321..759C}, for example, the
characteristic frequency at $\sim 0.01$ Hz in Cyg X-1 corresponds to
the viscous frequency ($\sim v_r/r$) of a geometrically thin accretion
flow at a truncation radius of about $25 R_{\rm S}$. X-ray timing data
typically contain several characteristic frequencies, while
theoretical models allow for different mechanisms of variability
(eg. \cite{2004A&A...427..251G}). The model presented here does not
contain enough physics to make predictions about the source and nature
of X-ray variability. For reference we note that, for a truncation
radius at $20 R_{\rm S}$, the viscous frequencies for our model are
$\sim 0.8$ Hz for the hot layer, $\sim 2$ Hz for the hot ring, and
$\sim 50$ Hz for the ADAF. 1 Hz is a characteristic frequency often
observed in Cyg X-1.

The physics of interaction of an ion-supported flow with a cool
disk is very well defined. The process produces a hot layer of tightly
constrained thickness and temperature. The physics is detailed enough,
for example, to be implemented in a numerical hydrodynamic simulations
of the accretion flow. The present model falls short of achieving
this: the inner hot ring where the evaporation into the ADAF takes
place, in particular, contains parametrizations that would need to be
improved with a more detailed (2-D) treatment of radiative
transfer. Thus the model used here has adjustable parameters but
compared with other hot corona models there is a straightforward path
to more rigorous calculations.

Note, however, that the likely presence of a strong ordered magnetic
field in the inner regions of the accretion (see 4.1 and 4.2 above)
adds additional physics that is not included either in existing hot
corona models or the present ion illumination model.

\section{Conclusions}\label{sec:conclusions}

From energetic considerations, the hard spectra observed in the
low-hard state of LMXBs must be produced by hot ($kT_{e}\sim$ 100 keV)
matter in the inner regions surrounding the black hole. If there is
also a much cooler disk present, there will necessarily be some degree
of interaction between the two components, and the disk will be
somewhat heated by irradiation from the hot Comptonizing component.
The fits reported in \cite{2006ApJ...652L.113M} and
\cite{2006ApJ...653..525M} neglect this interaction by fitting the
disk and hard component separately. In this paper we have shown that
incorporating the effects of this interaction heats the inner regions
of a moderately truncated disk so that, when coupled with the effects
of interstellar absorption, the size of the soft excess matches
observations. Our work also highlights the potential pitfalls of
using simple power law or analytic Comptonization fits at low
energies, which can provide significant deviations in the soft X-rays,
thus changing the shape and intensity of the observed soft excess.

In the case of GX 339-4, our model predicts an Fe-K component of
comparable strength to that observed, although we did not do a
detailed comparison. However, work by others has suggested that part
of the broadening in the Fe-K line that was observed for GX 339-4 can
be attributed to a large outflow, and detailed models of Fe-K
fluorescence in galactic black holes show lines that are much broader
than is found in AGN models (and which are normally used to fit
spectra). 

The model we have envisioned presents several opportunities for
further improvement, in order to better constrain the introduced
fitting parameters, $C$ and $\eta$. The spectrum from the hot ring
and ADAF are particularly uncertain, and dependent on a more detailed
model for the radiative transfer through this region, as well as the
source and number of seed photons (which will set the electron
temperature in both regions). The model's global accretion rate (which
is limited by the rate at which the hot layer spills over into the hot
ring and then evaporates into the ADAF) is also very low, although
this can be increased if the viscosity in the warm layer can be
increased, perhaps as a result of accretion through an ordered
magnetic field.

CD'A acknowledges financial support from the National Sciences and
Engineering Research Council of Canada. \bibliography{Black_hole}
\end{document}